\RequirePackage[T1]{fontenc}
\documentclass[12pt]{article}

\usepackage[height=8.85in,width=6.45in]{geometry}

\usepackage[utf8]{inputenc}
\usepackage{amsmath}
\usepackage{amssymb}
\usepackage{mathtools}
\numberwithin{equation}{section}
\usepackage{slashed}
\usepackage{braket}
\usepackage[svgnames]{xcolor}
\usepackage[colorlinks,citecolor=DarkGreen,linkcolor=FireBrick]{hyperref}
\usepackage{cite}
\usepackage{graphicx}
\usepackage{tikz}
\usepackage{tikz-cd}
\usepackage{times}
\usepackage{courier}
\usepackage{bm}
\usepackage{subfig}

\usepackage{xcolor}
\usepackage{mdframed}

\renewenvironment{figure}[1][]{
  \begin{originalfigure}[#1]
    \begin{mdframed}[linecolor=black!0,backgroundcolor=black!1]
}{
    \end{mdframed}
  \end{originalfigure}
}

\def\bZ{\mathbb{Z}}
\def\bR{\mathbb{R}}
\def\cA{\mathcal{A}}
\def\cC{\mathcal{C}}
\def\cO{\mathcal{O}}
\def\sT{\mathsf{T}}
\def\U{\mathrm{U}}
\def\id{\text{id}}
\def\RP{\mathbb{RP}}
\def\CP{\mathbb{CP}}

\def\unoriented{\text{unoriented}}
\def\Arf{\mathop{\text{Arf}}}
\def\Ker{\mathop{\text{Ker}}}
\def\Im{\mathop{\text{Im}}}
\def\Re{\mathop{\text{Re}}}
\def\vev#1{\langle#1\rangle}
\def\underline#1{\bm{#1}}
\begin{document}

\begin{titlepage}

\begin{flushright}
IPMU-18-0074
\end{flushright}

\vskip 3cm

\begin{center}

{\Large \bfseries A study of time reversal symmetry 
of abelian anyons}

\vskip 1cm
Yasunori Lee and Yuji Tachikawa
\vskip 1cm

\begin{tabular}{ll}
 & Kavli Institute for the Physics and Mathematics of the Universe, \\
& University of Tokyo,  Kashiwa, Chiba 277-8583, Japan
\end{tabular}

\vskip 1cm

\end{center}

\noindent
We perform a  study of  time reversal symmetry of abelian anyons $\mathcal{A}$  in 2+1 dimensions,
in the spin structure independent cases.
We will find the importance of the group $\mathcal{C}$ of \emph{time-reversal-symmetric anyons} modulo \emph{anyons composed from an anyon and its time reversal}.
Possible choices of local Kramers degeneracy are given by quadratic refinements of the braiding phases of $\mathcal{C}$, and the anomaly  is then given by  the Arf invariant of the chosen quadratic refinement.
We also give a concrete study of the cases when $|\mathcal{A}|$ is odd or $\mathcal{A}=(\mathbb{Z}_2)^N$.

\end{titlepage}

\setcounter{tocdepth}{2}
\tableofcontents


\section{Introduction and summary}
\paragraph{Motivations:}
Topological quantum field theories (TQFTs) in 2+1 dimensions have been studied 
for three decades from many points of view:
they not only have a natural place in the successful interaction of mathematics and high-energy physics,
but also give low-energy descriptions of two-dimensional gapped systems in condensed matter physics,
including the fractional quantum Hall materials and its generalizations. 

A relatively new theme in this field of study is how discrete symmetries act on such systems.
One novel aspect in this line of study is that sometimes these discrete symmetries can be anomalous,
in which case the system lives on the boundary of a symmetry-protected topological phase (SPT) in the bulk,
i.e.~the anomalous TQFT provides a gapped boundary of an SPT.

The basic formalism of symmetry actions on 2+1d TQFTs\footnote{%
In this paper, we restrict our analysis to  2+1d  \emph{non-spin} TQFTs, i.e.~those which do not require any specification of spin structures on the spacetime, unless otherwise explicitly stated.
This is mainly because no definitive reference on \emph{spin} 2+1d TQFTs and symmetry actions on them
have appeared in the literature.}
was laid out in \cite{Barkeshli:2014cna},
and a detailed analysis for the time-reversal symmetry was given in \cite{Barkeshli:2016mew}.
In these references one can find the entire formalism together with various interesting examples,
which mostly involved non-abelian anyons.

What we aim to provide in this paper is a study of time reversal actions on \emph{abelian} anyon systems.
Abelian anyons are far simpler than non-abelian anyons,
and the group $\bZ_2$ generated by the time reversal is far easier than the general group of symmetries.
This makes many of the necessarily complicated equations in \cite{Barkeshli:2014cna,Barkeshli:2016mew} more accessible.
Still, the TQFT structure in the abelian anyons show many of the features of general non-abelian anyons,
and we can hope that the anti-unitarity inherent in the time-reversal might give us an interesting twist in the analysis.

\paragraph{Some examples:}
A major source of abelian anyon systems is the abelian Chern-Simons theories,
whose action is given by $S= (i/2\pi) K_{IJ} \int a^I d a^J$, for $N$ $\U(1)$ gauge fields $a^I$, $(I=1,\ldots, N)$ and an integer matrix $K_{IJ}$.\footnote{%
Some of the examples mentioned below will need the spin structure to be specified on the spacetime to be well-defined,
but this subtlety does not play a role in the rough discussion in this introduction.
}.
The time reversal $\sT$ can act on $a^I$ by a matrix $\sT^I_J$ such that $\sT^2=1$.
Then the classical action is invariant under the time reversal if and only if $\sT K \sT=-K$.
Time-reversal-invariant abelian anyon systems in this setup was studied in detail in \cite{Chan:2015nea}.
One obvious example is the $\U(1)_k \times \U(1)_{-k}$ theory with the action $S=(i/2\pi) k\int (ada -bdb)$ such that $\sT$ exchanges $a$ and $b$.

There are however subtler examples of time-reversal-symmetric abelian anyon systems known in the literature.
One example is the so-called semion-fermion system, which is the $\U(1)_2\times \U(1)_{-1}$ theory
with the action $S=(i/\pi) \int ada - (i/2\pi) \int bdb$ \cite{Metlitski:2014xqa,Seiberg:2016rsg}.
In this case,  there is no integer matrix $\sT$ acting on $\U(1)$ gauge fields $a$ and $b$ such that $\sT K \sT= -K$, 
and therefore the time-reversal action cannot even be implemented at the level of this Lagrangian. 
One manifestation is that this time-reversal action has an anomaly.

\paragraph{Methods and objectives:}
In order to cover these subtler cases as well, we use an abstract formalism to describe abelian anyons.
We first have the finite abelian group $\cA$ of anyon charges,
such that for each anyon type $a\in \cA$ its topological spin $\theta(a)=e^{2\pi i h(a)} \in \U(1)$ is given.
We then consider an arbitrary time reversal action $\sT:\cA\to \cA$ such that $\sT^2=\id$,
with the only constraint that it reverses the spin, $\theta(\sT a)=\overline{\theta(a)}$.

Because of this generality, a general time-reversal symmetry $\sT$ can first have a \emph{symmetry localization obstruction} \cite{Barkeshli:2016mew,Barkeshli:2017rzd}.
When the obstruction is non-vanishing, the symmetry of the anyon system is not  $\bZ_2=\{\id,\sT\}$ but is a 2-group obtained by extending this $\bZ_2$ by the 1-form symmetry group $\cA$ \cite{Tachikawa:2017gyf,Benini:2018reh}.
When the obstruction vanishes, we can then study the \emph{anomaly of the time-reversal symmetry},
which is known to be characterized by two signs $Z_\text{anomaly}(\RP^4)=\pm1$ and $Z_\text{anomaly}(\CP^2)=\pm 1$,
which are the partition functions of the corresponding 3+1d SPT characterizing the anomaly.
In the following, we simply use the word \emph{obstruction} for the symmetry localization obstruction,
and the word \emph{anomaly} for the time-reversal anomaly.\footnote{%
This is not the standard usage in the literature, where they are often both called obstructions or anomalies.
Hopefully our usage is clearer.
}
In particular, there is a  formula \cite{Barkeshli:2016mew} \begin{equation}
Z_\text{anomaly}(\RP^4)=\frac{1}{|\cA|^{1/2}}\sum_{a=\sT a} \theta(a)\eta(a)\label{anomalyformulaX}
\end{equation} 
computing the anomaly from the topological spins $\theta(a)=\pm1$ and the \emph{local Kramers degeneracy}
$\eta(a)=\pm1$ for time-reversal-symmetric anyons $a=\sT a$: 
$\eta(a)$ can loosely be thought of as the local eigenvalue of $\sT^2$ associated to the anyon $a$.

Many natural questions then arise, for example: 
i) What are the allowed form of the time-reversal actions $\sT$ on abelian anyons $\cA$?
ii) Are there cases where the obstruction is non-vanishing?
iii) What can be said about the anomalies, assuming that the obstruction vanishes?
This paper is our first step toward answering these questions.

We will see the importance of the group $\cC$ defined as follows:
\begin{equation}
\cC=\{a= \sT a  \mid a\in \cA \} / \{ c+\sT c \mid c\in \cA \}.\label{defCintro}
\end{equation} 
In words, this is the group of \emph{time-reversal-symmetric anyons} modulo \emph{anyons composed from an anyon and its time reversal}.
When the obstruction vanishes, 
we will see that different allowed choices of the local Kramers degeneracy $\eta$ is classified by this group $\cC$.
Also, when the obstruction vanishes, 
we will see that  the anomaly \eqref{anomalyformulaX} can be rewritten as \begin{equation}
Z_\text{anomaly}(\RP^4)=\frac{1}{|\cC|^{1/2}} \sum_{[a]\in \cC} \theta(a)\eta(a),\label{above}
\end{equation} simplifying a sum over time-reversal-invariant anyons in $\cA$ into a sum over $\cC$.
We give a general analysis showing that $\cC=(\bZ_2)^{2m}$,
and the anomaly \eqref{above} is the associated Arf invariant.

We will also analyze  two classes of explicit examples in detail: one is when $|\cA|$ is odd,
and another is when $\cA=(\bZ_2)^N$.
Among them, we will not find any explicit example whose time-reversal symmetry is obstructed in this paper.

We also carry out a general analysis if it is possible or not to choose a linear function $\eta$ on time-reversal-invariant anyons valued in $\{\pm1\}$ such that the anomaly formula \eqref{anomalyformulaX} yields $\pm1$.
We will see that it is always possible to choose such an $\eta$.
Non-existence of such an assignment of $\eta$ was used as a sufficient condition for the existence of the obstruction in \cite{Barkeshli:2017rzd}.
Our analysis therefore says that at least with this simplified method we cannot find any abelian anyon system whose time-reversal anomaly is obstructed.

Finally, in a recent paper \cite{Benini:2018reh}, it was shown using the anomaly inflow that any unitary finite group symmetry on any abelian anyon system is not obstructed. 
These observations strongly suggest that the time-reversal symmetry of an abelian anyon system is never obstructed.
It would be interesting to further investigate if this conjecture holds or not.

\paragraph{Organization of the paper:}
The rest of the paper is organized as follows.
We start in section~\ref{sec:formalism} by reviewing the formalism we need.
We spell out the defining data of abelian anyons,
discuss three major sources of such systems,
and recall the Moore-Seiberg data associated to them.
We then explain how to express the time-reversal action in this formalism,
and how to compute the obstruction and the anomaly.

In section~\ref{sec:general}, 
we study what can be said about general time-reversal-symmetric abelian anyon systems,
without using the detailed features of the Moore-Seiberg data.
We will see that the anomaly formula \eqref{anomalyformulaX} can be re-written in terms of a sum over 
the group $\cC$ defined as in \eqref{above}, which is the  Arf invariant of $\theta(a)\eta(a)$ on $\cC$.

In section~\ref{sec:odd} and \ref{sec:even},
we consider concrete cases of time-reversal actions on abelian anyon systems.
As our preceding analysis will make clear, the situations differ drastically depending on 
whether an anyon can be divided by two or not.
We study two extreme cases.
Namely, in section~\ref{sec:odd} we study the case when $|\cA|$ is odd.
There, a straightforward argument shows that the theory is necessarily a gauge theory for an abelian group $A$ with a trivial time-reversal action on $A$, such that $|\cA|=|A|^2$.
The obstruction and the anomaly will vanish automatically.
In section~\ref{sec:even} we study the case when every element of $\cA$ is order two, i.e.~when $\cA=(\bZ_2)^N$.
There, a recent mathematical result allows us to enumerate all possible time-reversal actions.
We study the obstruction and the anomaly for each of these cases by direct computations using a computer program.
We will not find any case with obstructions.

\section{Basics of abelian anyons and their time reversal}
\label{sec:formalism}
\subsection{Defining data of abelian anyons}
Let us first review the defining data of abelian anyons in 2+1 dimensions.
In this paper we restrict to the case where the system is \emph{non-spin},
by which we mean that the system is well-defined without specifying the spin structure on the manifold.

Three main sources of such theories are  $\U(1)^N$ Chern-Simons theories,
non-abelian Chern-Simons theories at level 1,
and  gauge theories of finite abelian groups.
We prefer to use a democratic formalism which treat the output of these distinct methods equally.

Following \cite{Belov:2005ze,Stirling:2008bq,Kapustin:2010hk}, we take the defining data of a system of abelian anyons to be the triple $(\cA,\theta,c)$ where
\begin{itemize}
\item the group of charges of anyons $\cA$ is finite and abelian,
\item the topological spin $\theta:\cA\to \U(1)$ is a non-degenerate homogeneous quadratic function,
\item and the chiral central charge $c$  is an integer satisfying the Gauss sum constraint.
\end{itemize}
Here, a function $\theta:\cA\to \U(1)$ is called \emph{quadratic} if 
the braiding phase defined by\begin{equation}
B(a,b):=\theta(a+b) \theta(a)^{-1} \theta(b)^{-1}
\end{equation} is bilinear;
it is called \emph{non-degenerate} if $B$ is non-degenerate;
and it is called \emph{homogeneous} if \begin{equation}
\theta(na)=\theta(a)^{n^2}.
\end{equation}
The \emph{Gauss sum constraint} is the condition 
\begin{equation}
\sum_\cA \theta(a) = \sqrt{|\cA|} e^{2\pi i c/8}.\label{gauss}
\end{equation}
This constraint determines $c$ modulo 8.
As this description is rather abstract, let us discuss examples.

\subsection{Examples of abelian anyons}
\subsubsection{Abelian Chern-Simons}
The first set of examples are  $\U(1)^N$ Chern-Simons theories.
The action is given in Euclidean signature by the formula \begin{equation}
S=i\frac{K_{IJ}}{2\pi} \int_M a^{I} da^J 
\end{equation} where $a^I$ for $I=1,\ldots,N$ are $\U(1)$ gauge fields on the 3d manifold $M$
and $K_{IJ}$ is a symmetric matrix.
For this Lagrangian to be well-defined on 3d  oriented manifolds without specifying the spin structure,
$K_{IJ}$ needs to be an integral matrix such that the diagonal entries are even.
We call such matrix an even integral matrix.
The analysis of abelian Chern-Simons theory using the matrix $K$ is often called the $K$-matrix formalism in the condensed-matter literature.

The charges of the anyons are characterized by a finite abelian group \begin{equation}
\cA=\bZ^N / K\bZ^N,
\end{equation}
and an anyon $a=(a_I)\in \cA$ has the topological spin \begin{equation}
\theta(a)=e^{2\pi i\cdot \frac12 a_I (K^{-1})^{IJ} a_J}.
\end{equation} 
The right hand side is a well-defined function of $a+K\bZ^N$ thanks to the fact that $K$ is even and integral.
The braiding phase between two anyons $a,b\in \cA$ is given by \begin{equation}
B(a,b)=e^{2\pi i\cdot a K^{-1} b} 
\end{equation} 
which is bilinear, symmetric and non-degenerate.
The topological central charge $c$ of the system is the signature of $K$, namely the difference of the number of positive eigenvalues and the number of negative eigenvalues of $K$,
and it is a classic mathematical result that the Gauss sum constraint \eqref{gauss} is satisfied.

\subsubsection{Non-abelian Chern-Simons at level 1}
The second set of examples are non-abelian Chern-Simons theories $G_k$ with level $k$,
when $G$ is simply-laced and $k=1$.
In fact such a theory is equivalent to a $\U(1)^N$  Chern-Simons theory
where $N$ is the rank of $G$ and the associated $K_{IJ}$ defines the root lattice of $G$.

In particular, when $G=E_8$, the root lattice is equivalent to the weight lattice,
and the group of anyon is trivial, $\cA=0$.
This theory still has a nontrivial chiral central charge $c=8$.

\subsubsection{Finite group gauge theories}
\label{sec:fin}
The third set of examples are  gauge theories of finite Abelian group $A$.
For these theories, the group of anyons is $\cA=A\oplus \hat A$ where $\hat A$ is the Pontrjagin dual of $A$, 
namely the group of 1-dimensional representations of $A$.
Physically, an anyon $a\in A$ carries a magnetic flux specified by $a$
and an anyon $\chi\in \hat A$ carries an electric charge specified by $\chi$.
The topological spins are given by \begin{equation}
\theta(a+\chi)=\chi(a),
\end{equation} and the chiral central charge is zero.

This third set can be further generalized by introducing a nonzero Dijkgraaf-Witten action $\omega\in H^3(A,\U(1))$; 
the anyons become non-abelian in general, but they remain abelian when $\omega$ satisfies a certain simplifying condition \cite{Dijkgraaf:1989hb,Coste:2000tq}.
This simplifying condition is automatically met when $A=\bZ_n$. 

\subsubsection{Universality of abelian Chern-Simons constructions}
In \cite{Belov:2005ze} the quantization of $\U(1)^N$ Chern-Simons theories was analyzed carefully,
and two different Lagrangians leading to the same triple $(\cA,\theta,c)$ are shown to be dual, 
i.e.~are equivalent as quantum mechanical theories.
Conversely, it is a classic mathematical result \cite{Wall1,Wall2,Nikulin} that any triple $(\cA,\theta,c)$ comes from an even integral lattice $(\bZ^N,K_{IJ})$.
Here we note that the chiral central charge $c$ is determined by $\theta$ by the Gauss sum relation \eqref{gauss} mod 8,
and the mod 8 part can be freely changed by tensoring the $E_8$ level 1 theory or its orientation reversal.

This means that we do not lose any generality by assuming that the anyon system we consider in fact comes from an abelian Chern-Simons theory.
This point of view might be mentally reassuring to some of the readers.

\subsection{Moore-Seiberg data of abelian anyons}
As recalled above, an anyon system is characterized by the triple $(\cA,\theta,c)$.
In order to perform the computation as a 3d topological quantum field theory,
we need the Moore-Seiberg data \cite{Moore:1988qv,Moore:1989vd},
or equivalently, we need to describe anyons as a modular tensor category. 
The Moore-Seiberg data of abelian anyons forming a cyclic group $\bZ_n$ were discussed in Appendix E of \cite{Moore:1988qv};
the data for the general case were discussed in detail e.g.~in \cite{DongLepowsky,Stirling:2008bq}.
We quickly recall salient features below.

In general, a 3d topological quantum field theory is specified by morphisms \begin{equation}
F(a,b,c): a\otimes (b\otimes c) \stackrel{\sim}{\longrightarrow} (a\otimes b)\otimes c
\end{equation} describing the fusion and \begin{equation}
R(a,b): a\otimes b \stackrel{\sim}{\longrightarrow} b\otimes a
\end{equation} describing the half-braiding,
where $a$, $b$, $c$ are three arbitrary anyons,
satisfying the pentagon and hexagon relations.

\paragraph{Required relations and equivalences:}
For a system of abelian anyons, these morphisms $F(a,b,c)$ and $R(a,b)$ can be thought of simply as phases $\in \U(1)$.
Then the pentagon relation is \begin{equation}
F(a,b,c+d) F(a+b,c,d)=F(b,c,d) F(a,b+c,d) F(a,b,c)\label{pentagon}
\end{equation} and the hexagon relations are
\begin{equation}
\begin{aligned}
R(a,b+c)&=F(a,b,c){}^{-1} R(a,b) F(b,a,c) R(a,c) F(b,c,a){}^{-1},\\
R(a+b,c)&=F(a,b,c) R(b,c) F(a,c,b){}^{-1} R(a,c) F(c,a,b).
\end{aligned}\label{hexagon}
\end{equation}
The half-braiding $R(a,b)$ and the data $\theta(a)$, $B(a,b)$ are related by the formula \begin{equation}
\theta(a)=R(a,a),\qquad B(a,b)=R(b,a) R(a,b).\label{braidbraid}
\end{equation}

The phases $F(a,b,c)$ and $R(a,b)$ are not basis independent in the following sense.
For each pair of anyons $a,b$, we can introduce phases \begin{equation}
\U(1)\ni U(a,b): a\otimes b \stackrel{\sim}{\longrightarrow} a\otimes b
\end{equation} and we can define \begin{equation}
\begin{aligned}
(U.F)(a,b,c) &:=  U(b,c) U(a+b,c){}^{-1} U(a,b+c) U(a,b){}^{-1} F(a,b,c),\\
(U.R)(a,b) &:= U(a,b){}^{-1} U(b,a) R(a,b).
\end{aligned}
\label{FURU}
\end{equation}
The pairs $(F,R)$ and $(U.F,U.R)$ are considered physically equivalent. 
In particular, they correspond to the same $\theta$ and $B$.

There are $U(a,b)$ such that $(F,R)=(U.F,U.R)$.
This happens if and only if\footnote{%
The if part can be checked by a simple computation.
To show the only if part, suppose one is given such a $U$.
From $F=U.F$, $U$ is a two-cocycle, and determines an extension $0\to\U(1)\to \hat\cA\stackrel{p}{\to} \cA\to 0$,
such that for a section $s:\cA\to \hat\cA$ we have $s(a)s(b)=U(a,b)s(a+b)$.
From $R=U.R$, we see $U(a,b)=U(b,a)$.  Therefore $\hat\cA$ is Abelian.
We now construct another section $t:\cA\to \hat\cA$ as follows.
We pick an ordered basis, and choose $t(g_i)$ such that $p(t(g_i))=g_i$ and $t(g_i)^{n_i}=1$.
Then, for $a=\sum a_i g_i$, we define $t(a)=\prod_i t(g_i)^{a_i}$.
We can check $t(a)t(b)=t(a+b)$. 
We finally find $\beta(a)$ via the relation  $\beta(a)t(a)=s(a)$.
} there is a function $\beta: \cA\to \U(1)$ such that \begin{equation}
U(a,b)= \beta(a) \beta(b) /\beta(a+b)\label{d_A}.
\end{equation}

\paragraph{Existence and essential uniqueness:}
It is known that for any $\theta$ where $\theta$ is a homogeneous quadratic function on $\cA$,
there is a unique equivalence class of $(F,R)$. 
This can be seen as follows. 

First we show that an explicit representative can be constructed by giving an \emph{ordered basis} on $\cA$ (see e.g.~\cite{Quinn:1998un,Stirling:2008bq,Kapustin:2010hk}).
Namely, we fix a decomposition $\cA\simeq \bZ_{n_1}\times \bZ_{n_2}\times \cdots \times \bZ_{n_k}$,
and fix generators $g_i$ of $\bZ_{n_i}$.
We call such a choice of an ordered basis by $\cO$.
Then an arbitrary element $a\in \cA$ can be written as $a=\sum a_i g_i$, where $0\le a_i < n_i$.
We then define \begin{align}
(F_\cO)(a,b,c)&:=\prod_i \begin{cases}
1  & \text{if}\ b_i+c_i< n_i,\\
 \theta(g_i)^{n_i a_i} & \text{if}\ b_i + c_i \ge n_i
\end{cases}, \label{boo}\\
(R_\cO)(a,b) &:=\prod_i \theta(g_i)^{a_i b_i} \prod_{i<j} B(g_i,g_j)^{a_i b_j}.
\label{poo}
\end{align}
Next, given another pair $(F,R)$ for a given $(\cA,\theta)$ with an ordered basis,
there is an explicit algorithm given in Sec.~2.5 of \cite{Quinn:1998un}
which produces an appropriate $U$ such that $(U.F,U.R)=(F_\cO,R_\cO)$.

\paragraph{Other constructions of the Moore-Seiberg data:}
The data $(F,R)$ can also be given in terms of $K_{IJ}$ for the abelian Chern-Simons theory \cite[Chapter 12]{DongLepowsky}, as we describe below.
Since any finite abelian anyon system  comes from an abelian Chern-Simons theory as reviewed above,
this also provides the existence proof of the Moore-Seiberg data for arbitrary abelian anyon systems.

We denote by $\cA=\Lambda^*/\Lambda$ where $\Lambda^*=\bZ^N$, $\Lambda=K\bZ^N\subset \Lambda^*$.
We denote the inner product on $\Lambda^*$ by $\vev{\alpha,\beta}=\alpha_I (K^{-1})^{IJ} \beta_J$.
Note that its restriction makes $\Lambda$ an even integral lattice.

We now fix a function $\delta:L\times L\to \U(1)$ satisfying the following condition:
\begin{equation}
\frac{\delta(\alpha,\beta)}{\delta(\beta,\alpha)}=(-1)^{\vev{\alpha,\beta}} \qquad \text{if}\quad \alpha,\beta\in L_0.\label{cocycle-factor}
\end{equation}
This is the so-called cocycle factor, which also appears in the careful definition of the vertex operators of 2d chiral bosons on $\bR^n/\Lambda$, see e.g.~\cite[p.19]{Polchinski2}.
An example of such a $\delta$ is given by  \begin{equation}
\delta(\alpha,\beta):=e^{\pi i \sum_{I<J} (K^{-1}\alpha)^I K_{IJ} (K^{-1}\beta)^J},\label{example-delta}
\end{equation}
but any other choice will do.

For each anyon $a\in \cA$, we fix a lift $\alpha_a \in \Lambda^*$.
We then define \begin{equation}
R_\delta(a,b)=\frac{\delta(\alpha_a,\alpha_b)}{\delta(\alpha_b,\alpha_a)} e^{\pi i \vev{\alpha_a,\alpha_b}}
\end{equation} and \begin{equation}
F_\delta(a,b,c)=\frac{R_\delta(a+b,c)}{R_\delta(a,c)R_\delta(b,c)}.
\end{equation}
The pair $(F_\delta,R_\delta)$ defined above satisfies the required properties \eqref{pentagon}, \eqref{hexagon} and \eqref{braidbraid}.

There is yet another way to describe the Moore-Seiberg data of an abelian anyon system $(\cA,\theta)$,
which is intermediate between the one using the ordered basis and the one using the abelian Chern-Simons system.
This can be found in Appendix D of \cite{DGNO}.

\paragraph{Relation to the anomaly of 1-form symmetries:}
The discussion in the last paragraph establishes that the set of the equivalence classes of the pair $(F,R)$ is in one-to-one correspondence with the set of homogeneous possibly-degenerate quadratic functions on $\cA$.
This set is known to be equal to 
\begin{equation}
H^4(K(\cA,2),\U(1)) \label{anomaly}
\end{equation} from an old work of Eilenberg and Mac Lane \cite{EilenbergMacLane}.

More recently \cite{Kapustin:2014zva,Gaiotto:2014kfa}, 
it was noticed that the cohomology group $H^{d+2}(K(\cA,k+1),\U(1))$ 
 characterizes the anomaly of a $(d+1)$-dimensional system with the $k$-form symmetry $\cA$.
 Therefore, the object \eqref{anomaly} classifies the anomaly of $1$-form symmetry $\cA$ in 2+1 dimensions.
In our case, the point is to regard the group $\cA$ of abelian anyons as giving the 1-form symmetry of the system.
Then, the worldlines of abelian anyons labeled by elements of $\cA$ define 
a 1-cycle in $Z_1(M_3,\cA)$ which acts as the background gauge field for the 1-form symmetry $\cA$,
and the topological spins $\theta$ and the braiding phases $B$ describe the change in the phase of the partition function as we change the 1-cycle in $Z_1(M_3,\cA)$ keeping its homology class in $H_1(M_3,\cA)$.
Therefore, they describe the anomaly of the 1-form symmetry $\cA$.

\paragraph{$S$ and $T$ matrices:}
The discussions in this subsection up to this point did not require the non-degeneracy of $B$;
in particular, the identification of the equivalence classes of $(F,R)$ with \eqref{anomaly}
needs homogeneous quadratic functions $\theta$ which lead to degenerate $B$,
for example $\theta(a)\equiv 1$.
Therefore, the preceding discussions are more about the structure of the one-form symmetry $\cA$ in 2+1 dimensions.

For $(\cA,q)$ and the associated data $(F,R)$ to actually describe a topological quantum field theory,
we need the non-degeneracy of $B$.
In this case, the modular matrices are given by \begin{equation}
S_{ab}=\frac1{|\cA|^{1/2}}B(a,b),\qquad
T_{ab}=e^{-2\pi i c/24} \delta_{ab} \theta(a).
\label{ST}
\end{equation}
$S$ is invertible if and only if $B$ is non-degenerate: the non-degeneracy means that $B(a,b)$ is a character table of the abelian group $\cA$.

\subsection{The time reversal, the obstruction and the anomaly}

Group  actions on general, possibly non-abelian anyons were discussed in detail in \cite{Barkeshli:2014cna,Cui:2016wkw} from mathematical and condensed-matter points of view.
The equations discussed there were rather cumbersome.
Here we restrict our attention to the action of time reversal on abelian anyons.

\paragraph{Time reversal $\sT$ and associated objects $U$, $\beta$ and $\Omega$ :}
We denote the action of time-reversal on the anyons by \begin{equation}
\sT:\cA\to \cA
\end{equation} 
which we require to satisfy $\sT^2=\id$.
We require \begin{equation}
\theta(\sT a)=\overline{\theta(a)}.
\end{equation}

We fix the Moore-Seiberg data $(F,R)$ for $(\cA,\theta)$.
Let us now define the time-reversed Moore-Seiberg data $(\sT F,\sT R)$  by the formula \begin{equation}
\sT F(a,b,c):=\overline{F(\sT a,\sT b,\sT c)},\qquad
\sT R(a,b):=\overline{R(\sT a,\sT b)}.
\end{equation}
The pair $(\sT F,\sT R)$ also forms a Moore-Seiberg data for $(\cA,\theta)$.
Therefore, there are phases $U(a,b)$ such that \begin{equation}
(\sT F,\sT R)=(U.F,U.R).\label{defU}
\end{equation} where we remind the reader that the right hand side is defined in \eqref{FURU}.

Note that we trivially have $(F,R)=(\sT \sT F,\sT \sT R)$.
Computing the right hand side using \eqref{defU} twice,
we have
\begin{equation}
(F,R)=(\kappa.F,\kappa.R) 
\qquad
\text{where}
\qquad
\kappa(a,b):=  \overline{U(\sT a ,\sT b)}  U(a,b).
\end{equation}
Therefore,  there should be phases $\beta(a)$ as in \eqref{d_A} such that 
\begin{equation}
\overline{U(\sT a ,\sT b)}  U(a,b)=\beta(a)\beta(b)/\beta(a+b).\label{defbeta}
\end{equation} 
We now define \begin{equation}
\Omega(a):=\overline{\beta(\sT a)} / \beta(a).\label{defOmega}
\end{equation}
Using \eqref{defbeta}, one finds that $\Omega$ is linear, i.e.
$
\Omega(a+b)=\Omega(a)\Omega(b).
$

\paragraph{Choices in the construction:}
Recall that we started from $\sT$, from which we got $U$ via \eqref{defU}, from which we got $\beta$ via \eqref{defbeta}, from which we got $\Omega$ via \eqref{defOmega}.
There are certain indeterminacies at each stage.

If $U$ satisfies \eqref{defU}, \begin{equation}
\hat U(a,b)=U(a,b) \gamma(a)\gamma(b)/\gamma(a+b)\label{hatU}
\end{equation}
for any $\gamma$ also satisfies the same equation, as discussed around \eqref{d_A}.
Correspondingly, $\beta$ is changed but $\Omega$ is unchanged:\begin{equation}
\hat\beta(a):=\beta(a) \overline{\gamma(\sT a)}\gamma(a),
\qquad
\hat\Omega = \Omega.
\end{equation}

If $\beta(a)$ satisfies \eqref{defbeta},
\begin{equation}
\tilde \beta(a):=\beta(a) \nu(a) \label{betanu}
\end{equation}
 equally solves the same equation if $\nu$ is linear, i.e.~if $\nu(a+b)=\nu(a)\nu(b)$.
This changes 
$\Omega(a)$ to \begin{equation}
\tilde\Omega(a):=\Omega(a) \overline{\nu(\sT a)}/\nu(a).\label{tildeomega}
\end{equation}

\paragraph{The obstruction $[\Omega]$ :}
Now, we note that any linear map $f:\cA\to \U(1)$ is realized as $f(a)=B(a,\underline{f})$ for some $\underline{f}\in \cA$, 
since $B$ is assumed to be non-degenerate.
Therefore, $\Omega(a)$ corresponds to an element $\underline{\Omega}\in\cA$.
Similarly, $\nu$ appearing in \eqref{betanu} and \eqref{tildeomega} was also assumed to be linear, 
and therefore we have a corresponding element $\underline{\nu}\in \cA$, and we have \begin{equation}
\underline{\tilde \Omega}=\underline{\Omega} - \underline{\nu}+\sT \underline{\nu}.\label{equiv}
\end{equation}
Therefore, the choice-independent content is the equivalence class \begin{equation}
[\Omega]\in \cA/(1-\sT)\cA.
\end{equation}
We call this element  the \emph{obstruction}.
In other words, the obstruction vanishes $[\Omega]=0$ if and only if we can solve the following equation:\begin{equation}
\underline{\Omega}=\underline{\nu}-\sT \underline{\nu}.\label{omegaeq}
\end{equation} 
It is known that \cite{Tachikawa:2017gyf,Cordova:2018cvg,Benini:2018reh}
when the obstruction is non-vanishing,  the group $\bZ_2=\{1,\sT\}$  is not quite the group of symmetries of the system,
but rather is non-trivially extended by the 1-form symmetry $\cA$.

In passing, we mention that 
it is not at all clear whether the obstruction generally vanishes
in this description.
Some sub-cases when it vanishes can be established.
In Sec.~5.1.1 of \cite{Cui:2016wkw} and in the Appendix of \cite{EtingofGalindo},
the obstruction was shown to vanish when $|\cA|$ is odd.
Similarly, the obstruction can be shown to vanish when $|G|$ is odd.
Also, the obstruction obviously vanishes when one can find an abelian Chern-Simons realization
such that $\sT$ is actually an order-2 symmetry of $K_{IJ}$ which furthermore preserves the cocycle factor \eqref{cocycle-factor}.

\paragraph{The object $\eta$ :}
When the obstruction vanishes,
the group $G$ acts as a genuine symmetry.
In this case, there is a choice of $\beta(a)$ such that $\underline{\Omega}=0\in \cA$.
To emphasize that this is a special case, it is useful to denote such a choice of $\beta(a)$ by a different letter $\eta(a)$.
More explicitly, $\eta(a)$ needs to satisfy \begin{align}
\overline{U(\sT a ,\sT b)}  U(a,b)&=\eta(a)\eta(b)/\eta(a+b), &
\eta(a)\eta(\sT a)&=1.
\label{etadef}
\end{align}
We immediately see that
\begin{equation}
\eta(a)=\pm1, \qquad 
\eta(a+b)=\eta(a)\eta(b)\label{etaprop}
\end{equation}
if $\sT a=a$ and $\sT b =b$.
This quantity $\eta(a)$ for $\sT a=a$ has the interpretation of the local eigenvalue of $\sT^2$,
and sometimes called the local Kramers degeneracy \cite{Barkeshli:2016mew}.

Note that the choice of $\eta$ is  not unique.
We can  replace $\eta$ following \eqref{betanu} as follows:
 \begin{equation}
\tilde\eta(a):=\eta(a) \nu(a) = \eta(a) B(\underline{\nu},a).\label{etanu}
\end{equation}  
This solves \eqref{etadef}  if and only if \begin{equation}
\underline{\nu}= \sT \underline{\nu}.\label{ker}
\end{equation}

If we replace $U$ by $\hat U$ in \eqref{hatU},  
$\eta$ is replaced by \begin{equation}
\hat\eta(a) := \eta(a) \overline{\gamma(\sT a)}\gamma(a),\label{etagamma}
\end{equation} 
which satisfies the relations \eqref{etadef} automatically.
In particular, when $\gamma(a)=B(\underline{\gamma},a)$,
the change \eqref{etagamma} corresponds to the change \eqref{etanu} with \begin{equation}
\underline{\nu}=\underline{\gamma}+\sT\underline{\gamma}.\label{im}
\end{equation}
We physically identify two choices of $\eta$ different by this type of $\underline{\nu}$.
In other words, physical equivalence classes of allowed $\eta$ are parameterized by 
$\underline{\nu}$ satisfying \eqref{ker} modulo $\underline{\nu}$ given by \eqref{im};
two allowed $\eta$'s are different by an element in \begin{equation}
\cC:=\Ker(1-\sT)/\Im (1+\sT). \label{cC}
\end{equation} 
This is the group which we introduced in \eqref{defCintro} in the introduction.
Mathematically, we say that the set of $\eta$ is a torsor over $\cC$.

Note that changing $\eta$ using $\underline{\nu}$ does not change its value on $\Im(1+\sT)$,
as can be checked easily. 
In fact $\eta(c+\sT c)$ can be written in terms of $B$. 
To see this, one first sets $a=c$, $b=\sT c$ in the first equation of \eqref{etadef} to show \begin{equation}
\eta(c+\sT c)=U(\sT c, c)U(c,\sT c)^{-1},\label{zzz}
\end{equation} where we used the second equation of \eqref{etadef}.
Now, the explicit form of the equation $\sT R=U.R$  in \eqref{defU} is \begin{equation}
\overline{R(\sT a,\sT b)}=U(a,b)^{-1} U(b,a) R(a,b).
\end{equation}
Setting $a=c$, $b=\sT c$ again, we find \begin{equation}
U(\sT c, c)U(c,\sT c)^{-1}= R(c,\sT c)^{-1} R(\sT c, c)^{-1} =B(c,\sT c)^{-1}.
\end{equation}
Combining with \eqref{zzz}, we conclude that \begin{equation}
\eta(c+\sT c)B(c,\sT c)=1.\label{etaB}
\end{equation}

\paragraph{Anomalies :}
In general, the time reversal symmetry of non-spin $(d+1)$ dimensional systems are characterized by the phase given by the symmetry protected topological phase in $(d+2)$ dimensional unoriented spacetime,
which is a homomorphism to $\U(1)$ from the cobordism group $\Omega^\unoriented_{d+2}$ \cite{Kapustin:2014tfa,Kapustin:2014dxa,Freed:2016rqq,Yonekura:2018ufj}.
In our case $\Omega^\unoriented_4=\bZ_2\times \bZ_2$ is generated by $\RP^4$ and $\CP^2$,
and therefore the anomaly is characterized by two signs $Z_\text{anomaly}(\RP^4)$ and $Z_\text{anomaly}(\CP^2)$.

These two signs were computed in \cite{Barkeshli:2016mew} for general 3d non-spin topological quantum field theories;
see also \cite{Wang:2016qkb,Tachikawa:2016cha,Tachikawa:2016nmo} for the spin case.
For abelian anyons, the formulas of \cite{Barkeshli:2016mew} becomes \begin{equation}
Z_\text{anomaly}(\RP^4)=\frac{1}{|\cA|^{1/2}} \sum_{a = \sT a} \theta(a)\eta(a),\qquad
Z_\text{anomaly}(\CP^2)=\frac{1}{|\cA|^{1/2}} \sum_{a } \theta(a) = e^{2\pi i c/8}.
\label{anomalyformula}
\end{equation}
Since $Z_\text{anomaly}(\CP^2)$ is uniquely fixed in terms of $c$, we will only be interested in $Z_\text{anomaly}(\RP^4)$ below.

\section{Time reversal and the anomaly formula}
\label{sec:general}
\subsection{General properties to be established}
In this section, we study the property of the time reversal $\sT$ on general abelian anyon systems.
Since we do not have a good control of the Moore-Seiberg data $(F,R)$ in the general case,
we will use only the following information in this section, namely:
\begin{itemize}
\item The time reversal $\sT:\cA\to \cA $ satisfies $\sT^2=\id$,
\item $\theta(\sT a) =\theta(a)^{-1}$, and $B(a,b):=\theta(a+b)\theta(a)^{-1}\theta(b)^{-1}$ is non-degenerate,
\item $\eta(a)=\pm 1$  and $\eta(a+b)=\eta(a)\eta(b)$ if $a$ and $b$ are $\sT$ invariant, see \eqref{etaprop}.
\end{itemize}

\noindent We find the following general properties:
\begin{enumerate}
\item The group $\cC=\Ker(1-\sT)/\Im (1+\sT)$ introduced in \eqref{cC} is  $=(\bZ_2)^n$ for some $n$.
\item The non-degenerate pairing $B:\cA\times\cA\to \U(1) $ restricts to a non-degenerate pairing on $B:\cC\times \cC\to \{\pm1\}$.
Furthermore, $n$ is even, $n=2m$.
\item  There is an obstruction if the summand $q(a):=\theta(a)\eta(a)$ on $\Ker (1-\sT)$ does not restrict to a function on $\cC$. 
\item There is always a choice of $\eta(a)$ such that $q(a)=\theta(a)\eta(a)$ restricts to a function on $\cC$.
\item If $q(a)=\theta(a)\eta(a)$ restricts to a function on $\cC$, then the anomaly $Z_\text{anomaly}(\RP^4)$ is the Arf invariant of $q:\cC\to \{\pm1\}$.
In particular, there are $2^{m-1}(2^m+1)$ choices of $\eta$'s for which the anomaly vanishes,
and $2^{m-1}(2^m-1)$ choices of $\eta$'s for which the anomaly is non-vanishing.
\end{enumerate}

In \cite{Barkeshli:2017rzd},
the non-existence of the assignment $\eta$ so that $Z_\text{anomaly}(\RP^4)=\pm1$ was considered as a simple sufficient condition to see if a group action on anyon systems is obstructed.
Our Property 4 here means that this simplified method does not allow us to find any obstructed time-reversal action in the case of abelian anyons.

\subsection{Derivations of the properties}
Let us show these properties. 
\paragraph{Property 1 :}
We start from a  trivial observation that \begin{equation}
\Im(1+\sT) \subset \Ker (1-\sT)
\label{eee}
\end{equation} which simply follows from $\sT^2=1$.
We now consider the group \begin{equation}
\cC=\Ker (1-\sT)/ \Im(1+\sT),
\end{equation} i.e.~the group of time-reversal invariant anyons modulo anyons which are composites of an anyon and its time reversal.
Every element in $\cC$ is order two, since $2a=a+\sT a$ if $(1-\sT)a=0$.
Therefore $\cC=(\bZ_2)^n$ for some $n$.

\paragraph{Property 2 :}

We first note that \begin{equation}
B(a,\sT b)=B(\sT a,b)^{-1}.
\end{equation} 
Therefore, we have \begin{equation}
B((1+\sT)a,b)=B(a,(1-\sT)b).
\end{equation}
Therefore \begin{equation}
\Ker (1-\sT) \subset [\Im (1+\sT)]^\perp,\qquad
\Ker (1+\sT) \subset [\Im (1-\sT)]^\perp.
\label{ccc}
\end{equation}
Using the non-degeneracy of $B$, we have \begin{equation}
|\Ker (1-\sT) | \le |\cA|/|\Im (1+\sT)|,\qquad
|\Ker (1+\sT) | \le |\cA|/|\Im (1-\sT)|
\label{aaa}
\end{equation}
Now, we obviously have \begin{equation}
|\cA/\Ker (1+\sT)|=|\Im (1+\sT)|, \qquad
|\cA/\Ker (1-\sT)|=|\Im (1-\sT)|.
\label{bbb}
\end{equation}
The relations \eqref{aaa} and \eqref{bbb} together shows that the inclusions in \eqref{ccc} are actually   equalities: \begin{equation}
\Ker (1-\sT) = [\Im (1+\sT)]^\perp,\qquad
\Ker (1+\sT) = [\Im (1-\sT)]^\perp.
\label{ddd}
\end{equation}
The relation \eqref{eee} means that  $B(a,b)$ on $\cA$ descends to  bilinear forms on $\cC$
and the relation \eqref{ddd} means that $B$ thus defined on $\cC$ are actually non-degenerate.

We now recall the standard fact that any non-degenerate pairing $B$ on $(\bZ_2)^n$ can be put into either of the following two forms:\footnote{%
The proof goes as follows, see e.g.~\cite[Theorem 2.1]{Dugger}. 
Consider a non-degenerate pairing $B$ on a finite-dimensional $\bZ_2$-vector space $V$.
As a zeroth step, we note that  any $B(a,b)=\pm 1$, because $B(a,b)^2=B(2a,b)=1$.
Then, as a first step, we show that $V$ is a direct sum of an orthogonal part and a symplectic part.
To see this,
if there is a element $x\in V$ such that $B(x,x)=-1$, one takes the orthogonal complement of $x$,
and repeat the process.
Eventually, there is no $x\in V$ such that $B(x,x)=-1$.
Then, pick a nonzero $x\in V$ randomly. From non-degeneracy, there is a $y\in V$ such that $B(x,y)=-1$.
Then we take the orthogonal complement of $x$ and $y$, and repeat the process.
As a second step, one shows that $V_3=(\bZ_2)^3$ with the orthogonal $B$ can in fact be split into a one-dimensional orthogonal vector space plus a two-dimensional symplectic space. 
This can be done by taking the orthogonal complement of the vector $(1,1,1)\in V_3$.
This completes the proof.
}
\begin{itemize}
\item Symplectic: there is a basis $u_1,v_1,$, $u_2,v_2$, \ldots, $u_m,v_m \in \cA$ with $n=2m$  such that \begin{equation}
B(u_i,u_j)=B(v_i,v_j)=1,\qquad B(u_i,v_j)=\begin{cases}
1 & (i\neq j)\\
-1 & (i=j)
\end{cases}.\label{sympl}
\end{equation}
\item Orthogonal: there is a basis $u_1,\ldots,u_n\in \cA$ such that \begin{equation}
B(u_i,u_j)=\begin{cases}
1 & (i\neq j)\\
-1 & (i=j)
\end{cases}.\label{ortho}
\end{equation} 
When $n$ is odd, the orthogonal complement of the vector $(1,1,\ldots,1)$ has a symplectic structure as given above.
\end{itemize}

In our case, $B$ on $\cC$ satisfies $B(a,a)=+1$ for all $a\in \cC$,
since if we regard $a\in \Ker(1-\sT)$, $B(a,a)=\theta(2a)/\theta(a)^2=\theta(a)^2=1$;
the last equality follows since $\theta(a)=\theta(\sT a)=\overline{\theta(a)}$.
Therefore, $B$ should be of the symplectic type, so that $n$ is even: $n=2m$.
Therefore \begin{equation}
|\cA|= |\Im(1+\sT)|^2 \cdot 2^{2m}.
\end{equation}

\paragraph{Property 3 :}
Let us now try to evaluate the anomaly \eqref{anomalyformula} \begin{equation}
Z_\text{anomaly}(\RP^4)=\frac{1}{|\cA|^{1/2}} \sum_{a = \sT a} \theta(a)\eta(a)
= \frac{1}{|\Im (1+\sT)| 2^m} \sum_{a\in \Ker(1-\sT) } \theta(a)\eta(a).
\end{equation}
Let us perform the sum over $\Ker(1-\sT)$ by first summing within a coset  $a'\in a+\Im (1+\sT)$
and then over $\cC$.
We would like to relate, then, $\theta(a+c+\sT c)\eta(a+c+\sT c)$ and $ \theta(a)\eta(a) $. 
We have\footnote{%
Note that in this section we are analyzing the anomaly without actually using the Moore-Seiberg data.
It is still useful to recall that we saw in \eqref{etaB} that $B(c,\sT c) \eta(c+\sT c)=1$
when the obstruction can be shown to vanish using the Moore-Seiberg data.
} \begin{equation}
\frac{\theta(a+c+\sT c)\eta(a+c+\sT c)}{\theta(a)\eta(a)}=   B(c,\sT c) \eta(c+\sT c).\label{etaobs}
\end{equation} 
Here we note that $B(c+c',\sT (c+c'))=B(c,\sT c)B(c',\sT c')$, and $B(c,\sT c)=\pm1$.
Therefore, the right hand side is a homomorphism $\Im (1+\sT)\to \{\pm1\}$.
If this is non-trivial, the sum over $a'\in a+\Im(1+\sT)$ simply vanishes, and we have \begin{equation}
Z_\text{anomaly}(\RP^4)=0,
\end{equation} which should not happen if there is no obstruction.

\paragraph{Property 4 :}
Resuming the discussion,
let us ask if we can choose an $\eta$ such that the right hand side of \eqref{etaobs} is a constant $=1$.
For this purpose, we regard $f(c):=B(c,\sT c)$ as a one-dimensional representation of $\Im(1+\sT)$.

Since $\Im(1+\sT)$ is a subgroup of $\Ker(1-\sT)$,
this $f$ can be extended (non-uniquely) to a one-dimensional representation of $f:\Ker(1-\sT)\to \U(1)$.
Let $q(a)=\theta(a)f(a)$ for $a\in \Ker (1-\sT)$.
By construction, this function on $\Ker(1-\sT)$ is constant on a coset $a'\in a+\Im(1+\sT)$,
and therefore restricts to a function $\cC\to \U(1)$ satisfying $q(a+b)=q(a)q(b) B(a,b)$ on $\cC$.
Since $B$ on $\cC$ is of symplectic type, $q(a)=\pm1$.\footnote{%
This can be shown by actually constructing one quadratic refinement $q$ for $B$ in the standard form.
This turns out to take value in $\{\pm1\}$.
Every other $q$ is obtained by multiplying it by a homomorphism $\cC\to \U(1)$ which is necessarily valued in $\{\pm1\}$, the statement follows.}
Since $\theta(a)=\pm1$, we conclude that $f(a)=\pm1$.
Therefore we can use this $f(a)$ on $\Ker(1-\sT)$ as $\eta(a)$, since
$f$ satisfies every condition which should be satisfied by $\eta$.

\paragraph{Property 5 :}
As discussed, without an obstruction, $\theta(a')\eta(a')$ should be constant on $a'\in a+\Im(1+\sT)$.
Let us denote this function by $q(a)=\theta(a)\eta(a): \cC\to \bZ_2$.
Then we have \begin{equation}
Z_\text{anomaly}(\RP^4,\eta)= \frac{1}{|\cC|^{1/2}} \sum_{a\in \cC } q(a).\label{000}
\end{equation} 

We note that the function $q(a)$ satisfies $B(a,b)=q(a+b)q(a)^{-1}q(b)^{-1}$ on $\cC$.
Therefore, $q$ is a non-degenerate homogeneous quadratic function on $\cC\simeq (\bZ_2)^{2m}$.
Such a function $q$ is known as a quadratic refinement of $B$,
and for such a $q$, the right hand side of \eqref{000} is known as its Arf invariant, which is known to take values in $\{\pm1\}$\footnote{%
Another place where the Arf invariant appears is in the description of the spin structure on a Riemann surface \cite{Atiyah,Johnson}. 
Briefly, for a Riemann surface $\Sigma$, we let $\cC:=H^1(\Sigma,\bZ_2)$,
and $B(a,b)=\int_\Sigma a \cup b$ for $a,b\in \cC$.
This $B$ is non-degenerate, and moreover, $B(a,a)=+1$.
We define $q:\cC\to \bZ_2$ so that $q(a)$ is $+1$ / $-1$  if the spin structure is Neveu-Schwarz / Ramond around a non-intersecting loop representing the Poincar\'e dual to $a$, respectively.
This function $q$ is known to satisfy $q(a+b)=q(a)q(b)B(a,b)$,
and its Arf invariant is defined as the right hand side of \eqref{000}.
The spin structure is called even or odd depending on whether the Arf invariant is $+1$ or $-1$. 
}: \begin{equation}
Z_\text{anomaly}(\RP^4,\eta) = \Arf q.
\end{equation}

It is a standard result that for $\cC=(\bZ_2)^{2m}$ 
there are $2^{m-1}(2^m+1)$ choices of $q$'s for which the Arf invariant is $+1$,
and there are $2^{m-1}(2^m-1)$ choices of $q$'s for which the Arf invariant is $-1$.
This in particular means that if the obstruction vanishes, there is at least one assignment of $\eta$
which makes the system free of the time-reversal anomaly $Z_\text{anomaly}(\RP^4)$.

\subsection{A comment on the spin case}
\label{sec:spin}
At this level of generality, it is not difficult to extend the analysis to abelian anyon systems which depends on the spin structure.
The main difference is that among the anyons there is a special anyon, sometimes called the \emph{transparent} fermion $f\in \cA$ such that $2f=0$ and $\theta(f)=-1$.
Anyons in $\cA$ such that $B(f,a)=1$ are in the Neveu-Schwarz sector $\cA_\text{NS}$,
while those with $B(f,a)=-1$ are in the Ramond sector $\cA_\text{R}$.
Then, for any anyon $a\in \cA_\text{NS}$, we have $af\in \cA_\text{NS}$ and $\theta(af)=-\theta(a)$.
It is known that $\eta(af)=-\eta(f)$.

Since $a$ and $af$ always appear in pairs, we consider physically distinct anyons in the Neveu-Schwarz sector to be  labeled by $\cA':=\cA_\text{NS}/\{0,f\}$;
this is why $f$ is called \emph{transparent}.
The braiding $B$ descends to a non-degenerate pairing on $\cA'$,
and we then require that the time reversal $\sT$ to be an order-2 operation on $\cA'$.

For  systems which do not feel the spin structure, the anomaly $Z_\text{anomaly}(\RP^4)=\pm1$ as we reviewed above.
For systems which do feel the spin structure, the anomaly is in general given by \begin{equation}
Z_\text{anomaly}(\RP^4) = e^{2\pi i \nu /16}\label{16}
\end{equation} for an integer $\nu$ modulo 16.
A generalization of the anomaly formula  for this was found in \cite{Wang:2016qkb,Tachikawa:2016nmo} 
and is given for abelian anyons by \begin{equation}
Z_\text{anomaly}(\RP^4)=\frac{1}{|\cA'|^{1/2}} \sum_{a = \sT a } \theta(a)\eta(a).
\end{equation} 
Now the sum is over time-reversal invariant anyons in $\cA'$.

Our analysis for the non-spin case can be repeated up to Property 3 without any change except the replacement of  $\cA$ by $\cA'$ everywhere.
The only  additional change in Property 5 is that $q(a)=\theta(a)\eta(a)$ is now a function $q:\cC\to \{\pm1,\pm i\}$,
and we still have \eqref{000}.
The right hand side in this case is known as the Brown-Arf invariant $=e^{2\pi i k/8}$ for an integer $k$ modulo 8.\footnote{%
The Brown-Arf invariant appears in the description of the pin$^-$ structure on a possibly non-orientable Riemann surface, see 
e.g.~Sec.~3 of \cite{KirbyTaylorPin} or the appendix of \cite{KlausThesis}.
The sum \eqref{000} is also a special case of the general Gauss sum \eqref{gauss}, since $(\cC,q)$ satisfies all the mathematical conditions to be an ordinary abelian anyon system.
}
Comparing with \eqref{16}, we conclude $\nu=2k$, that is, we found that the time-reversal anomaly of abelian anyons is always an even  integer modulo 16.

\section{Case study I : when $|\cA|$ is odd}
\label{sec:odd}
Let us discuss the case when $|\cA|$ is odd. 
We first determine the standard form of the time reversal action $\sT:\cA\to\cA$, which satisfies $\sT^2=\id$.
A greatly simplifying feature is that when $|\cA|$ is odd, one can always divide an anyon by two, 
in the sense that for any $a\in \cA$ there is a unique $b$ such that $a=2b$. 
We denote such this element $b$ by $a/2$. 
We note  that $(\sT a)/2=\sT (a/2)$.

Then any $a\in \cA$ is a sum $a=a_+ + a_-$ such that $\sT a_\pm = \pm a_\pm$, since we can explicitly take \begin{equation}
a_\pm = (a\pm \sT a)/2.
\end{equation}
This means that $\cA$ is a direct sum $\cA=\cA_+\oplus \cA_-$.

Now let us determine $B$  on $\cA$ compatible with this action of $\sT$.
For elements $a_+, b_+\in \cA_+$,  let $c_+=a_+/2$. Then we have \begin{equation}
B(c_{+},b_{+})=B({\sf T}c_{+},{\sf T}b_{+})^{-1}=B(c_{+},b_{+})^{-1}
\end{equation} 
and therefore $B(a_+,b_+)=1$.
We similarly have $B(a_-,b_-)=1$ for arbitrary $a_-,b_-\in \cA_-$.
In order for the braiding $B$ to be non-degenerate on $\cA=\cA_+\oplus \cA_-$,
this means that the non-trivial pairing should happen between $\cA_+$ and $\cA_-$.
Equivalently, $\cA_+ = \widehat{\cA_-}$, 
where $\hat G$ for an abelian group $G$ denotes its Pontrjagin dual.
Therefore we have $B(a_+,b_-)=a_+(b_-)$, where $a_+$ is now regarded as a homomorphism $\cA_-\to \U(1)$.

A compatible $\theta$ on $\cA$ is then determined as follows:
for any $a_\pm \in \cA_\pm$, we take $c_\pm=a_\pm/2 \in \cA_\pm$.
We have \begin{equation}
\theta(c_\pm)=\theta(\sT c_\pm)^{-1}=\theta(\pm c_\pm)^{-1}=\theta(c_\pm)^{-1}
\end{equation}
and therefore $\theta(a_\pm)=1$.
From this, we easily conclude that \begin{equation}
\theta(a_+ + b_-) = a_+(b_-).
\end{equation}

Comparing with the discussion in Sec.~\ref{sec:fin}, we find that this is a gauge theory with finite abelian gauge group $A=\cA_+$ with a trivial action of the time-reversal.
The anyons labeled by $\hat A= \cA_-$ are Wilson lines. 
For consistency,  an anyon $a\in \hat A=\cA_-$ is sent to $-a$ by the time reversal $\sT$.

We can easily see that there is no obstruction and there is no anomaly.
To see that there is no obstruction, we pick an ordered basis in $A=\cA_+$ and then a corresponding ordered basis in $\hat A=\cA_-$.
From the explicit formulas \eqref{boo} and \eqref{poo} of the Moore-Seiberg data $(F_\cO,R_\cO)$ in this ordered basis, we see that \begin{equation}
(\sT F_\cO, \sT R_\cO)  = (F_\cO,R_\cO).
\end{equation}
Then $U$ can be taken to be identically 1, and therefore the obstruction vanishes.
Then $\eta(a)$ is a linear function on $\cA$ which is $=\pm1$ if $a=\sT a$, i.e.~if $a\in A=\cA_+$.
This is identically $=+1$ since any $a$ is divisible by 2.
By the anomaly formula, we see that $Z_\text{anomaly}(\RP^4)=+1$, 
i.e.~the system is non-anomalous. 

The group $\cC$ is trivial. This alone allows us to conclude that the anomaly vanishes, using our general analysis given in the last section.

\section{Case study I\hspace{-.1em}I : $\cA=(\bZ_2)^N$}
\label{sec:even}
In the previous section we studied the case where $|\cA|$ was odd.
What made the analysis straightforward was that we can always divide an anyon by two.
In this section we consider the opposite extreme case, where any anyon $a\in\cA$ satisfies $2a=0$.
This means that $\cA\simeq (\bZ_2)^N$.

\subsection{Classification of the time reversal action $\sT$}
We first study all possible actions of $\sT$, compatible with the braid pairing $B$.
This was recently carried out with a different motivation in \cite{Dugger},
whose results we summarize below.

Recall that any non-degenerate pairing $B$ on $\cA=(\bZ_2)^N$ is given either by a symplectic one or an orthogonal one, as we discussed around \eqref{sympl}, \eqref{ortho}.
Second, all possible forms of $\sT$ were classified for both symplectic and orthogonal $B$ in \cite{Dugger}.
Note that, since $B(a,b)=\pm 1$, the condition for $\sT$ is that $B(\sT a,\sT b)=B(a,b)^{-1}=B(a,b)$.

When $B$ is symplectic, $\sT$ with respect to the standard basis is a direct sum of the three matrices: 
\begin{equation}
I=\left[\begin{array}{cc}
			1 & 0\\
			0 & 1
		\end{array}\right] , 
		\qquad
J=\left[\begin{array}{cc}
		0 & 1\\
		1 & 0
		\end{array}\right] , 
		\qquad
M=\left[\begin{array}{cccc}
		1 & 0 & 1 & 1\\
		0 & 1 & 1 & 1\\
		1 & 1 & 1 & 0\\
		1 & 1 & 0 & 1
		\end{array}\right].
\end{equation}

When $B$ is orthogonal, the situation is more complicated.
Again, we always choose the standard basis as in \eqref{ortho}.
When $N$ is odd, any $\sT$ action fixing $B$ is known to fix the vector $\omega=(1,\cdots,1)$,
i.e. $\sT\omega=\omega$.
This is because $\omega$ is uniquely characterized by the condition $B(a,a)=B(a,\omega)$ for all $a\in \cA$.
The orthogonal complement to $\omega$ carries a symplectic structure, and $\sT$ on it is given by the direct sum of $I$, $J$ and $M$ as above.

Let us move on to the case when $N=2n$ is even.
An important operation is a $\bZ_2$ operation on $n\times n$ matrices with entries $\in \bZ_2$, 
defined by \begin{equation}
m(M)_{ij}=1-M_{ij}.
\end{equation}
One can show that when $M^2=1$, $m(M)^2=1$. 
This operation $m$ is called the \emph{mirror} in \cite{Dugger}.
Any $\sT$ can then be conjugated to exactly one of the following forms: 
\begin{equation}
\begin{array}{ccc}
		I^{\oplus(n-k)}\oplus J^{\oplus k} & \text{and their mirrors} & (1\leq k \leq n-1),\\
		\\
		m\big(I^{\oplus(n-k)}\big)\oplus J^{\oplus k} & \text{and their mirrors} & (0\leq k \leq n-1),\\
		\\
		m\big(I^{\oplus(n-k-1)}\oplus J^{\oplus k}\big)\oplus J & & (1\leq k \leq n-2).\\
	\end{array}
\end{equation}
The last cases are conjugate to mirrors of their own.

Clearly, there is no need to study $\sT$ which is given by a direct sum. 
Therefore, we simply need to study the following cases:
\begin{align}
I,\quad J,\quad m(I^{\oplus (n-k)}\oplus J^{\oplus k}),\quad
m(m(I^{\oplus (n-k)})\oplus J^{\oplus k}).\label{choices}
\end{align}
We will see below that there are no compatible $\theta$ 
for $m(J^{\oplus l})$ with odd $l$
and $m(m(I^{\oplus n-k})\oplus J^{\oplus k})$ with even $k$.

We will tabulate the results  we obtained by  explicit computations in the following subsections. 
The computations are done as follows.
We start from $\cA$ with a specified $B$ and $\sT$. 
We fix an ordered basis $g_1,\ldots,g_n$ of $\cA$.
We first classify all compatible $\theta$. 
For each $\theta$, we compute $(F,R)$ in the standard basis.
We then find $U$ which satisfies $(\sT F,\sT R)=(U.F, U.R)$ using the algorithm given in Sec.~2.5 of \cite{Quinn:1998un}.
We then find one choice of $\beta$ by solving \eqref{defbeta}, which can be done by setting $\beta(g_i)=1$ for all basis elements
and finding $\beta(a)$ for linear combinations; 
it is guaranteed that there is such a $\beta$.
From this $\beta$ we can easily compute $\Omega$, and the obstruction is checked by whether we can solve \eqref{omegaeq}.
We implemented the algorithm we explained above in a computer program,
which is available upon request to the authors.

\subsection{When the braid pairing $B$ is symplectic}
We start by analyzing the symplectic cases.
\paragraph{$\blacksquare\ \sT=I$ :}
Our anyons are $\cA=\{0,u,v,u+v\}$ with the pairing $B(u,v)=-1$, $B(u,u)=B(v,v)=+1$.
There are four choices of $\theta:\cA\to \U(1)$.
Up to the relabeling of anyons, we can choose either of the two cases: 
\begin{equation}
\begin{aligned}
\text{a)}\qquad & \theta(u)=\theta(v)=1,\quad \theta(u+v)=-1,\\
\text{b)}\qquad & \theta(u)=\theta(v)=\theta(u+v)=-1.
\end{aligned}
\label{twochoices}
\end{equation}
The former corresponds to the standard $\bZ_2$ gauge theory,
also known as the toric code theory, and is the $\U(1)^2$ Chern-Simons theory with level matrix $\begin{pmatrix}
0 & 2\\
2 & 0
\end{pmatrix}$ whose action is $S= (i/\pi) \int adb$.

By an explicit computation, we find that the obstruction vanishes for both choices of $\theta$,
and $\eta$ is simply a linear function $\cA\to \U(1)$.
There are four choices of $\eta$ for each case.
For the case (a), three choices give $Z_\text{anomaly}(\RP^4)=+1$ and one choice $Z_\text{anomaly}(\RP^4)=-1$;
this one choice is when $\eta(u)=\eta(v)=-1$. 
This last case is sometimes called the \emph{eTmT} phase in the condensed-matter literature.

For the case (b) too, three choices give $Z_\text{anomaly}(\RP^4)=+1$ and one choice $Z_\text{anomaly}(\RP^4)=-1$;
this choice is when $\eta(u)=\eta(v)=+1$. 
The cases (a) and (b) can be distinguished by looking at $Z_\text{anomaly}(\CP^2)=+1$ for (a) and $Z_\text{anomaly}(\CP^2)=-1$ for (b).

\paragraph{$\blacksquare\ \sT=J$ :}
Our anyons are still $\cA=\{0,u,v,u+v\}$ with the pairing $B(u,v)=-1$, $B(u,u)=B(v,v)=+1$.
There are two choices of $\theta:\cA\to \U(1)$ compatible with $u=\sT v$,
which is again given by \eqref{twochoices}.
By an explicit computation, we find that the obstruction vanishes for both choices of $\theta$,
and $\eta(u+v)$ is forced to be $-1$.
Time-reversal invariant anyons are $0$ and $u+v$, and one finds $Z_\text{anomaly}(\RP^4)=+1$.

\paragraph{$\blacksquare\ \sT=M$ :}
Anyons are generated by $u_{1,2}$ and $v_{1,2}$.
Up to relabeling $u_1\leftrightarrow u_2$ and $v_1\leftrightarrow v_2$,
there is only one allowed choice of $\theta$, given by \begin{equation}
\theta(u_1)=+1,\quad
\theta(u_2)=-1,\quad
\theta(v_1)=+1,\quad
\theta(v_2)=-1.
\end{equation}
The time-reversal-invariant anyons are generated by $u_1+u_2$ and $v_1+v_2$,
and the group $\cC$ is trivial.

An explicit computation shows that the obstruction vanishes,
and there is only one allowed choice of $\eta$ which is $\eta(u_1+u_2)=\eta(v_1+v_2)=+1$.
One finds that $Z_\text{anomaly}(\RP^4)=+1$.

\subsection{When the braid pairing $B$ is orthogonal}
Let us move on to the case where $B$ is orthogonal.
We start by analyzing a few simple cases.

\subsubsection{Some simple cases}

\paragraph{$\blacksquare\ \cA=\bZ_2$ :}
We will start with the simplest case when $\cA=\bZ_2=\{0,a\}$, $B(a,a)=-1$. 
This means $\theta(a)=\pm i$. 
The time reversal action is $\sT a =a$, and therefore we cannot have $\theta(a)=\overline{\theta(\sT a)}$.
Therefore this is inconsistent as a non-spin theory.\footnote{%
As a spin theory this is consistent as we discussed in Sec.~\ref{sec:spin},
and describes the semion-fermion system,
which has $Z_\text{anomaly}(\RP^4)=e^{\pm 2\pi i/8}$.
}
Since $\sT=I$ is just two copies of this system, this is also inconsistent.

\paragraph{$\blacksquare\ \sT=J$ :}
Our anyons are $\cA=\{0,u,v,u+v\}$ with the pairing $B(u,u)=B(v,v)=-1$ and $B(u,v)=+1$.
Up to the relabeling of anyons, there is only one choice of $\theta$ compatible with this $\sT$:
\begin{equation}
\theta(u)=+i,\qquad \theta(v)=-i.
\end{equation}
Time-reversal-invariant anyons are $0$ and $u+v$, with the group $\cC$ being trivial.
By an explicit computation, we find that the obstruction vanishes,
and $\eta(u+v)=-1$.
We find $Z_\text{anomaly}(\RP^4)=+1$.

\paragraph{$\blacksquare\ \sT=m(I^{\oplus 2})$ :}
Anyons are generated by $u_{1,2,3,4}$.
Up to relabeling, there are three allowed choices of $\theta$, given by
\begin{equation}
\begin{aligned}
\theta(u_1)=+i,\quad \theta(u_2)=+i,\quad \theta(u_3)=+i,\quad \theta(u_4)=+i;\\
\theta(u_1)=+i,\quad \theta(u_2)=+i,\quad \theta(u_3)=-i,\quad \theta(u_4)=-i;\\
\theta(u_1)=-i,\quad \theta(u_2)=-i,\quad \theta(u_3)=-i,\quad \theta(u_4)=-i.
\end{aligned}
\end{equation}

The time-reversal-invariant anyons are generated by $u_1+u_2$, $u_1+u_3$ and $u_1+u_4$,
and the group $\cC=(\bZ_2)^2 $. An explicit computation shows that the obstruction vanishes,
and there exist various allowed choices of $\eta$.
One finds that both $Z_\text{anomaly}(\RP^4)=+1$ and $Z_\text{anomaly}(\RP^4)=-1$ may occur.

\paragraph{$\blacksquare\ \sT=m(I\oplus J)$ :}
Anyons are generated by $u_{1,2}$ and $v_{1,2}$.
Up to relabeling, there are four allowed choices of $\theta$, given by
\begin{equation}
\begin{aligned}
\theta(u_1)=+i,\quad \theta(u_2)=+i,\quad \theta(v_1)=+i,\quad \theta(v_2)=+i;\\
\theta(u_1)=+i,\quad \theta(u_2)=+i,\quad \theta(v_1)=-i,\quad \theta(v_2)=-i;\\
\theta(u_1)=-i,\quad \theta(u_2)=-i,\quad \theta(v_1)=+i,\quad \theta(v_2)=+i;\\
\theta(u_1)=-i,\quad \theta(u_2)=-i,\quad \theta(v_1)=-i,\quad \theta(v_2)=-i.\\
\end{aligned}
\end{equation}

The time-reversal-invariant anyons are generated by $u_1+u_2$ and $v_1+v_2$,
and the group $\cC$ is trivial. An explicit computation shows that the obstruction vanishes,
and there exist various allowed choices of $\eta$.
One finds that $Z_\text{anomaly}(\RP^4)=+1$.

\paragraph{$\blacksquare\ \sT=m(J^{\oplus 2})$ :}
Completely the same argument as $\sT=m(I\oplus J)$ case goes through, and one finds that $Z_\text{anomaly}(\RP^4)=+1$.

\subsubsection{The general case}

As already mentioned, 
we implemented the algorithm in a program and studied all the choices \eqref{choices}
from smaller $n$ to larger $n$.
We did not find any case where the time-reversal symmetry $\sT$ is obstructed.
This leads us to suspect that the time-reversal symmetry $\sT:\cA\to \cA$ on an abelian anyon system might be always un-obstructed. 
It would be interesting to study if this is the case or not.

Below, we study various choices of $\sT$ in \eqref{choices} using the anomaly formula.
As we will see, for some choice of $\sT$ there is no compatible $\theta$.

\paragraph{$\blacksquare\ \sT=m(I^{\oplus k})$ :}
Anyons are generated by $u_1,\cdots,u_{2k}$. 
A compatible $\theta$ is
\begin{equation}
\begin{array}{cc}
\theta(u_l) = +i & (l=1,\cdots,2k)
\end{array}
\end{equation}
for even $k$, and
\begin{equation}
\theta(u_l) = \left[\begin{array}{cl}
+i & (l\neq 2k)\\
-i & (l=2k)
\end{array}\right.
\end{equation}
for odd $k$.

$\sT$-invariant anyons consist  of the sum of even number of anyons,
since $\sT u_{i}\neq u_{i}$ and  $\sT(u_i+u_j)=u_i+u_j$.
The image of $1+\sT$ are either $0$ or $u_1+\cdots u_{2k}$.
Therefore $\cC=(\bZ_2)^{2k-2}$.

Let us see what the anomaly formula tells us.
$\eta$ is a $\{\pm1\}$-valued linear function on $\sT$-invariant anyons.
Let us arbitrarily extend it to a $\{\pm1\}$-valued linear function on the entire $\cA$.
We denote the extension by $\tilde \eta$.
We then have $\tilde\eta(u_l)\theta(u_l)=\pm i$,
and furthermore, $\tilde\eta(a)\theta(a)$ is either real or purely imaginary, depending on whether
$a$ is $\sT$-invariant or not.
Therefore we have
\begin{align}
Z_\text{anomaly}(\RP^4)
&=\frac{1}{\sqrt{2^{2k}}}\sum_{a\in\cA} \Re \tilde\eta(a)\theta(a) \nonumber \\
&=\frac{1}{2^{k}}\Re\Big[\big(1+ i\big)^{m}\big(1-i\big)^{2k-m}\Big]\nonumber \\
&= \left[\begin{array}{cl}
+1 & \big(k-m\equiv0\text{ mod }4\big),\\
0 & \big(k-m\equiv\pm1\text{ mod }4\big),\\
-1 & \big(k-m\equiv2\text{ mod }4\big)
\end{array}\right.
\label{bosh}
\end{align}
where $m$ is the number of the basis anyon $u_l$ such that $\tilde\eta(u_l)\theta(u_l)=+i$.

Each distinct choice of $\eta$ on $\sT$-invariant anyons corresponds to two choices of $\tilde\eta$ on the entire $\cA$.
Therefore, by a short computation from \eqref{bosh}, 
we see that there are $2^{k-2}(2^{k-1}\pm1)$ choices of $\eta$ for which the anomaly formula gives $\pm1$.
This agrees with our general discussion in Sec.~\ref{sec:general}, see Property 5.

\paragraph{$\blacksquare\ \sT=m(J^{\oplus k})$ :}
Anyons are generated by $u_1,v_1,\cdots,u_k,v_k$.
Odd $k$ is inconsistent since there is no compatible choice of $\theta$.
This can be seen as follows. We start from
\begin{equation}
\theta(\sT u_l)=\theta(v_l)^{-1}\prod_{m=1}^{k}\theta(u_m)\theta(v_m).
\end{equation}
Noting that both $\theta(u_l)$ and $\theta(v_l)$ are $\pm i$, we see
\begin{equation}
\prod_{m=1}^{k}\theta(u_m)\theta(v_m)=-\theta(u_l)\theta(v_l).
\end{equation}
We now introduce $c_m:=\theta(u_m)\theta(v_m)=\theta(u_m+v_m)=\pm 1$, where the last equality follows since $u_m+v_m$ is $\sT$-invariant.
We then have $\prod_{m=1}^k c_m = -c_l$ for arbitrary $l$. 
Therefore, we get $(\pm1)^k=\mp1$, which runs into a contradiction when $k$ is odd.

When $k$ is even, a compatible $\theta$ is \begin{equation}
\begin{array}{cc}
\theta(u_l) = \theta(v_l) = +i & (l=1,\cdots,k).
\end{array}
\end{equation}

A short computation shows that $\sT$-invariant anyons are the linear combinations of $u_i+v_i$'s,
and $u_i+v_i$ itself is in the image of $1+\sT$. 
Therefore the group $\cC$ is trivial.
The anomaly formula can be evaluated explicitly, using the signs $\eta(u_l+v_l)\theta(u_l+v_l)=\pm1$ for $l=1,\ldots,k$:
\begin{equation}
Z_\text{anomaly}(\RP^4)
=\frac{1}{2^{k}}\Big[\big(1+ 1\big)^{m}\big(1-1\big)^{k-m}\Big]
= \left[\begin{array}{cl}
+1 & \big(k=m\big),\\
0 & \big(k\neq m\big).
\end{array}\right.
\end{equation}
Indeed, there is only a single allowed choice of $\eta$, which is simply given by $\eta(u_l+v_l)=\theta(u_l+v_l)$.

\paragraph{$\blacksquare\ \sT=m(I^{\oplus(n-k)}\oplus J^{\oplus k})$ :}
Anyons are generated by $u_1,\cdots,u_{2(n-k)}$ and $w_1,x_1,\cdots,w_k,x_k$.
An allowed  $\theta$ is
\begin{equation}
	\begin{array}{cc}
		\theta(u_l) = +i & (l=1,\cdots,2(n-k)),
	\end{array}
   \quad
	\begin{array}{cc}
		\theta(w_m) = \theta(x_m) = +i & (m=1,\cdots,k)
	\end{array}
\end{equation}
for even $n$ and
\begin{equation}
	\theta(u_l) =\left[\begin{array}{cl}
		+i & (l\neq 2(n-k)),\\
      -i & (l=2(n-k)),
	\end{array}\right.
   \quad
	\begin{array}{cc}
		\theta(w_m) = \theta(x_m) = +i & (m=1,\cdots,k)
	\end{array}
\end{equation}
for odd $n$.
We do not repeat the analysis of the anomaly formula, since it is analogous to the ones we have already given above.

\paragraph{$\blacksquare\ \sT=m\big(m(I^{\oplus(n-k)})\oplus J^{\oplus k}\big)$ :}
Anyons are generated by $u_1,\cdots,u_{2(n-k)}$ and $w_1,x_1,\cdots,w_k,x_k$.
This time, even $k$ is inconsistent, since there is no compatible $\theta$. This can be seen as follows.
We start from
\begin{equation}
\theta(\sT u_l)=\theta(u_l)\cdot\prod_{m=1}^{k}\theta(w_m)\theta(x_m)
\end{equation}
Using $\theta(u_l)=\pm i$, we have
\begin{equation}
\prod_{m=1}^{k}\theta(w_m)\theta(x_m)=-1.\label{k}
\end{equation}
Next, we consider
\begin{equation}
\theta(\sT w_j)=\theta(x_j)^{-1}\prod_{m=1}^{k}\theta(w_m)\theta(x_m)\cdot\prod_{l=1}^{2(n-k)}\theta(u_l)
\end{equation}
which implies
\begin{equation}
\prod_{l=1}^{2(n-k)}\theta(u_l)=\theta(w_j)\theta(x_j),
\end{equation} for arbitrary $j$.
In other words, $\theta(w_j)\theta(x_j)=\pm1$ for arbitrary $j$, and this sign is independent of $j$.
This contradicts with \eqref{k} if $k$ is even.

For  odd $k$, a compatible $\theta$ is
\begin{equation}
	\begin{array}{cc}
		\theta(u_l) = +i & (l=1,\cdots,2(n-k)),
	\end{array}
   \quad
	\begin{array}{cc}
		\theta(w_m) = \theta(x_m) = +i & (m=1,\cdots,k)
	\end{array}
\end{equation}
for even $n$, and
\begin{equation}
	\theta(u_l) =\left[\begin{array}{cl}
		+i & (l\neq 2(n-k))\\
      -i & (l=2(n-k))
	\end{array}\right.,
   \quad
	\begin{array}{cc}
		\theta(w_m) = \theta(x_m) = +i & (m=1,\cdots,k)
	\end{array}
\end{equation}
for odd $n$.
Again, we do not repeat the analysis of the anomaly formula.

\section*{Acknowledgements}
The authors thank Francesco Benini and Po-Shen Hsin for explaining the content of \cite{Benini:2018reh}.
Y.L.~is partially supported by the Programs for Leading Graduate Schools, MEXT, Japan, via the Leading Graduate Course for Frontiers of Mathematical Sciences and Physics. 
Y.T.~is partially supported  by JSPS KAKENHI Grant-in-Aid (Wakate-A), No.17H04837 
and JSPS KAKENHI Grant-in-Aid (Kiban-S), No.16H06335,
and also by WPI Initiative, MEXT, Japan at IPMU, the University of Tokyo.

\bibliographystyle{ytphys}
\baselineskip=.95\baselineskip
\bibliography{ref}

\end{document}